\documentclass[11pt]{article}

\usepackage{amsmath}
\usepackage{amssymb}
\usepackage{times}
\usepackage{dsfont}
\usepackage{theorem}

\title{Origin of the Complex Structure of Quantum Mechanics}
\author{Andreas Walter Aste$^{a,b}$\\
$\quad$\\
$^{a}$\emph{Department of Physics, University of Basel, 4056 Basel, Switzerland}\\
$^{b}$\emph{Paul Scherrer Institute, Forschungsstrasse 111, 5232 Villigen PSI, Switzerland}}
\date{Mai 21, 2019}

\begin{document}
\maketitle

\begin{abstract}
This paper addresses the question why quantum mechanics is formulated in a unitary Hilbert space, i.e.
in a manifestly complex setting. Investigating the linear dynamics of real quantum theory in a
finite-dimensional Euclidean Hilbert space hints at the emergence of a complex structure.
A widespread misconception concerning the measurement process in quantum mechanics and
the hermiticity of observables is briefly discussed.
\end{abstract}
\maketitle
\section{Introduction}
It is a generally accepted postulate of quantum theory that the states of a quantum system are
represented by rays in a unitary, i.e. {\emph{complex}} (and separable) Hilbert space.
Whereas it is rather a matter of taste whether one wants to work in a complete
vector space or not - since all incomplete pre-Hilbert spaces can be completed to Hilbert spaces - the
question why nature seems to prefer a complex setting turned out to be a difficult one.\\

Sol\`er \cite{Soler} established that Quantum Mechanics may be formulated in real, complex or quaternionic Hilbert spaces
only. St\"uckelberg \cite{Stueckelberg1,Stueckelberg2} provided some physical but mathematically non-rigorous
arguments for ruling out a real Hilbert space formulation of quantum mechanics, assuming that any formulation should
incorporate a statement of a Heisenberg principle. Recently, Moretti and Oppio \cite{MorettiOppio} linked the complex
Hilbert space structure to the presence of Poincar\'e symmetry.\\

Of course, it is always possible to decompose a complex theory in some more or less elegant way into
real and imaginary parts to enforce a real picture of a quantum theory, but then the original structure of
the algebra of observables is lost and such a naive point of view does not explain why the
complex setting is so successful. In this paper, a simplified finite-dimensional view on the topic using linear algebra on an
undergraduate level only is presented in order to provide a simplified access to the subject also for students.
A concise discussion linking the measurement process in quantum mechanics to the hermitian character of
observables is provided.

\section{Real Quantum Mechanics in a Finite-Dimensional Hilbert Space}
For notational convenience, we use the fact that every finite dimensional real Hilbert space $\mathcal{H}_R^n$ is
isomorphic to a vector space $V_{\scriptscriptstyle{\mathds{R}}}^n = (\mathds{R}^n_{\scriptscriptstyle{\mathds{R}}}, \langle \cdot , \cdot  \rangle_R )$
with $n \in \mathds{N}$ and the standard Euclidean scalar product
\begin{equation}
\langle \cdot  , \cdot \rangle_R : V_{\scriptscriptstyle{\mathds{R}}}^n \times V_{\scriptscriptstyle{\mathds{R}}}^n  \rightarrow \mathds{R}\, , \quad
( \psi , \phi) \mapsto \langle \psi , \phi \rangle_R = \sum \limits_{k=1}^n \psi_k \phi_k = \psi^T \phi
\end{equation}
for all vectors
\begin{equation}
\psi =
\left(\begin{array}{c} \! 
\psi_1 \\  \psi_2 \\ \vdots \\ \psi_n  
\end{array} \right) \, , \quad
\phi =
\left(\begin{array}{c} \! 
\phi_1 \\  \phi_2 \\ \vdots \\ \phi_n  
\end{array} \right) \, ,
\end{equation}
in order to represent all physical state vectors in $\mathcal{H}_R^n$ directly by column vectors in
$V_{\scriptscriptstyle{\mathds{R}}}^n$. The row vector $\psi^T = (\psi_1, \ldots , \psi_n)$ denotes the transposed
vector of $\psi$ which allows to express the scalar product $\langle \psi, \phi \rangle = \psi^T \phi$
as a matrix multiplication.\\

Due to the linear structure of quantum mechanics,
a physical {\emph{state}} $\Psi$  is characterized by the {\emph{relative size}} of the components
$\psi_1, \psi_2, \ldots , \psi_n$ of a corresponding state {\emph{vector}} $\psi$, therefore a real state $\Psi$ can be represented
by a {\emph{normalized} state vector $\psi$ (or $-\psi$) with $|| \psi ||^2 = \langle \psi , \psi \rangle =1$.\\

If the quantum mechanical system under consideration  possesses a continuous one-parameter family of special orthogonal
symmetry transformations $R(\tau) \in SO(n) $,
\begin{equation}
SO(n) = \{ R \in GL(n, \mathds{R}) \mid R^T = R^{-1} , \,  \det R = 1 \} \, , 
\end{equation}
where $\tau$ can be interpreted as a real time parameter,
the real Schr\"odinger equation reads
\begin{equation}
\frac{d}{d \tau} \psi (\tau) = A \psi (\tau)
\end{equation}
with real skew-symmetric matrix $A$ generating special orthogonal transformations, hence $A$ is an element
of the $n(n-1)/2$-dimensional real Lie algebra $so(n)$
\begin{equation}
A \in so(n) \, , \quad so(n) = \{ A \in Mat(n, \mathds{R})  \mid A^T=-A \} \, .
\end{equation}
The real Schr\"odinger equation could also be written in the standard form
\begin{equation}
i \frac{d}{d \tau} \psi (\tau) = (iA) \psi (\tau) = H \psi (\tau)  \label{SchroedingerReal}
\end{equation}
with a hermitian, but purely imaginary pseudo-Hamilton operator $H = i A$ with vanishing diagonal elements.
Accordingly, all eigenvalues of $A$ are imaginary and real eigenvectors of $A$
do not exist in $V_{\scriptscriptstyle{\mathds{R}}}^n$.
If $\lambda = i \omega$ with $\omega \in \mathds{R}$ is an eigenvalue of $A$, $\lambda^* = -\lambda = -i \omega$
is also an eigenvalue, since for a corresponding complex eigenvector $v_\lambda$ where $A v_\lambda = i \omega v_\lambda$
one has the complex conjugate equation
\begin{equation}
A^* v^*_\lambda = A v^*_\lambda =  (i \omega)^* v^*_\lambda = -i \omega v^*_\lambda = -\lambda v^*_\lambda \, .
\end{equation}

The real solution of equation (\ref{SchroedingerReal}) is
\begin{equation}
\psi ( \tau) = e^{A \tau} \psi (0) = e^{-i H \tau} \psi (0) \, . \label{SchroRealSol}
\end{equation}

Now spectral theory tells us that a real skew-symmetric $n \times n$-matrix $A$ can be brought into
$2 \times 2$ block diagonal form $A'$ by a special orthogonal transformation $S \in SO(n)$
\begin{equation}
A' = S^{-1} A S = 
\left(\begin{array}{cccccccccc}
0 & \omega_1 &  &  &  &  &  &  &  &  \\
-\omega_1 & 0 &  &  &  &  &  &  &  & \\
& & 0 & \omega_2 & & & 0 &  &  &  \\
& &- \omega_2 & 0 & &  &  &  &   & \\
& & & &  \ddots &  &   &  &  &  \\
& & & & &  0 & \omega_r   & &   & \\
&  & 0 &   &   & -\omega_r &  0  &  &  &\\
&  & &  &   &   &    &  0 &  & \\
& &  &  &   & &   &   & \ddots & \\
&  &  &  &   &  &   &   &  & 0  \\
\end{array} \right)
\label{blockdiagonalA}
\end{equation}
with $\{ \omega_1, \ldots , \omega_r \} \subset \mathds{R}$,
where the solution (\ref{SchroRealSol}) becomes
\begin{equation}
\psi(\tau) = e^{S A' S^{-1} \tau} \psi (0) = S e^{A' \tau} S^{-1} \psi (0) 
\end{equation}
with
\begin{equation}
e^{A' \tau} = 
\left(\begin{array}{cccccccccc}
\cos (\omega_1 \tau) &  \sin (\omega_1 \tau)  &  &  &  &  &  &  &  &  \\
-\sin (\omega_1 \tau) & \cos ( \omega_1 \tau) &  &  &  &  & \mkern-80mu 0 &   &  & \\
& & \ddots & & & & &  &  &  \\
& & & & &  \cos (\omega_r \tau) & \sin (\omega_r \tau)   & &   & \\
&   \mkern-80mu 0 &  &   &   & -\sin (\omega_r \tau)&  \cos (\omega_r \tau)  &  &  &\\
&  & &  &   &   &    &  1 &  & \\
& &  &  &   & &   &   & \ddots & \\
&  &  &  &   &  &   &   &  & 1  \\
\end{array} \right) \, .
\label{blockdiagonalO}
\end{equation}
Therefore, the state vectors of any real finite-dimensional quantum mechanical system can be described
as an orthogonal superposition of $n-2r$ static states and $r$ pairs of mutually coupled oscillators.
The time evolution of such a pair in a two-dimensional subspace of $V^n_\mathds{R}$,
which can be represented by a coordinate vector in two-dimensional
Euclidean space in the block-diagonalizing basis implicitly introduced above, is given for $j=1, \ldots r$ by
\begin{equation}
\frac{d}{d \tau}
\left(\begin{array}{c}
\psi'_{j,1} (\tau ) \\
\psi'_{j,2} (\tau ) \\
\end{array} \right) =
\left(\begin{array}{cc}
0 & \omega_j   \\
- \omega_j  & 0 \\
\end{array} \right)
\left(\begin{array}{c}
\psi'_{j,1} (\tau ) \\
\psi'_{j,2} (\tau )\\
\end{array} \right)
\end{equation}
with the corresponding oscillation frequency $\omega_j$ and is solved by
\begin{equation}
\left(\begin{array}{c}
\psi'_{j,1} (\tau) \\
\psi'_{j,2} (\tau) \\
\end{array} \right) =
\left(\begin{array}{c}
\psi'_{j,1} (0) \cos ( \omega_j \tau) + \psi'_{j,2} (0) \sin (\omega_j \tau) \\
-\psi'_{j,1} (0) \sin ( \omega_j \tau) + \psi'_{j,2} (0)  \cos (\omega_j \tau)\\
\end{array} \right) \, . \label{RealSol}
\end{equation}
Equation (\ref{RealSol}) is completely analogous to the complex representation
\begin{displaymath}
\psi'_{j,1} (\tau)  + i \psi'_{j,2} (\tau) =
e^{-i \omega_j \tau} (\psi'_{j,1} (0)  + i \psi'_{j,2} (0) ) =
( \cos ( \omega_j \tau)  -i \sin ( \omega_j \tau ))  (\psi'_{j,1} (0) + i \psi'_{j,2} (0) )
\end{displaymath}
\begin{equation}
= ( \cos ( \omega_j \tau) \psi'_{j,1} (0) + \sin (\omega_j \tau) \psi'_{j,2} (0)) -
i ( \sin (\omega_j \tau) \psi'_{j,1} (0) - \cos (\omega_j \tau) \psi'_{j,2} (0) ) \, .
\label{RepComp}
\end{equation}
Discarding the physically irrelevant static sector in the real Hilbert space above,
the orthogonal dynamics of the remaining real $2r$-dimensional quantum system becomes equivalent to the
unitary dynamics of an $r$-dimensional system in a complex setting.\\

Of course, more concrete information is needed to characterize the frequency spectrum
$\{ \omega_1 , \ldots , \omega_r \}$ and the underlying physics of a quantum mechanical system.
Without additional physical conditions, the frequencies $\omega_{1, \ldots , r}$ can be positive or negative as well.
The truly diagonal part in the propagator (\ref{blockdiagonalO}) could be associated with static orthogonal state pairs
with vanishing frequencies $\omega_{r+1} , \ldots , \omega_{n/2}$ if $n=2m$ is even; for $n=2m+1$ odd a single static
unpaired state always remains.\\

The considerations so far are rooted in the fact that the unitary groups
\begin{equation}
U(n)=\{ U \in GL(n, \mathds{C}) \mid U^\dagger U = U^{*T} U =1 \}
\end{equation}
have maximal tori $T(U(n))$ which are homeomorphic to $n$-tori $T^n$, i.e. the $n$-fold topological products of circles $S^1$ 
\begin{equation}
T(U(n)) = \{ diag ( e^{i \varphi_1}, e^{i \varphi_2} , \ldots , e^{i \varphi_n}) \mid   \forall j   \, \, \varphi_j \in \mathds{R} \}
\simeq S^1 \times \ldots \times S^1 \simeq T^n \, ,
\end{equation}
and that the same is true for maximal tori of the orthogonal groups
\begin{equation}
T(SO(2m)) \simeq T(SO(2m+1)) \simeq T^m \, .
\end{equation}

However, whereas in ordinary quantum mechanics the expression $\psi'_{j,1} (\tau)  + i \psi'_{j,2} (\tau)$ in equation
(\ref{RepComp}) corresponds to a single stationary state, the elongations $(\psi'_{j,1} (\tau) , \psi'_{j,2} (\tau))$ rather
describe an oscillating superposition of two linearly independent state vectors in the real setting. This is not necessarily
a problem, since in a relative state interpretation of quantum mechanics the wave functions discussed above
are universal wave functions, and all of physics is presumed to follow from these functions alone - even the measurement
process itself.

%%%%%%%%%%%%%%%%%%%%%%%%%%%%%%%%%%%%%%%%%%%%%%%%%
%%%%%%%%%%%%%%%%%%%%%%%%%%%%%%%%%%%%%%%%%%%%%%%%%

\section{Real Three-Dimensional Quantum Mechanics}
As an exotic example quantum mechanics in three real dimensions - i.e. in a three-dimensional state space
$\mathcal{H}^3_R$ and {\emph{not}} in three-dimensional configuration space - shall be examined briefly in this section.\\

It is instructive to consider the complex three-dimensional case of a spin-1 system
in the Hilbert space $\mathcal{H}^3_C := \mathds{C}_{\mathds{C}}^3$ first.
Notationally simple generators $i \Sigma_{1,2,3} \in so(3)$ of the real Lie group $SO(3)$ which generate rotations around
the spatial $x^{1,2,3}$-axes are given by
\begin{equation}
\Sigma_1=\left(\begin{array}{ccc}
0 & 0 & 0 \\
0 & 0 & -i \\
0 & +i & 0 
\end{array}\right)\; , \quad
\Sigma_2=\left(\begin{array}{ccc}
0 & 0 & +i \\
0 & 0 & 0 \\
-i & 0 & 0 
\end{array}\right)\; , \quad
\Sigma_3=\left(\begin{array}{ccc}
0 & -i & 0 \\
+i & 0 & 0 \\
0 & 0 & 0 
\end{array}\right)  \, . \label{so3_gen_num1}
\end{equation}
The skew-symmetric operators above contain imaginary elements only in order to have the hermiticity
required for observables.
They satisfy the standard angular momentum $so(3)$ commutation relations for symmetry reasons
\begin{equation}
[\Sigma_l,\Sigma_m]= i \epsilon_{lmn} \Sigma_n \, .
\end{equation}
Alternatively one could use, e.g.,  the generators
$\Sigma'_1 = \Sigma_1$, $\Sigma'_2=\Sigma_3$, and $\Sigma'_3=-\Sigma_2$
which also fulfill the commutation relations $[\Sigma'_l,\Sigma'_m]= i \epsilon_{lmn} \Sigma'_n$.
It is common usage in physics to
treat spin-1 systems by working with an alternative operator basis
\begin{equation}
s_1= \frac{1}{\sqrt{2}} \left(\begin{array}{ccc}
0 & 1 & 0 \\
1 & 0 & 1 \\
0 & 1 & 0 
\end{array}\right) \, , \, \,
s_2= \frac{i}{\sqrt{2}} \left(\begin{array}{ccc}
0 & -1 & 0 \\
1 & 0 & -1 \\
0 & 1 & 0 
\end{array}\right) \,  , \, \,
s_3= \left(\begin{array}{ccc}
1 & 0 & 0 \\
0 & 0 & 0 \\
0 & 0 & -1 
\end{array}\right) \, . \label{so3_gen_num2}
\end{equation}
These spin operators are obtained from the change of basis
\begin{equation}
s_k= B^{-1} \Sigma'_k B \, , \quad 
B = \left(\begin{array}{ccc}
\frac{1}{\sqrt{2}} & 0 & -\frac{1}{\sqrt{2}} \\
0 & 1 & 0 \\
\frac{i}{\sqrt{2}} & 0 & \frac{i}{\sqrt{2}} 
\end{array}\right) \, , \quad
B^{-1} = \left(\begin{array}{ccc}
\frac{1}{\sqrt{2}} & 0 & -\frac{i}{\sqrt{2}} \\
0 & 1 & 0 \\
-\frac{1}{\sqrt{2}} & 0 & -\frac{i}{\sqrt{2}} 
\end{array}\right) \, .
\end{equation}

The exemplary definitions (\ref{so3_gen_num1}) and (\ref{so3_gen_num2}) exhibit
a central problem of three-dimensional real quantum physics: The skew-symmetric generators
$\Sigma_{1,2,3}$ (or the real $i \Sigma_{1,2,3})$ lead to vanishing expectation values on every real state vector
in $\mathcal{H}^3_R$.
On the other hand, the operator basis $s_{1,2,3}$ contains real {\emph{symmetric}} operators
$s_1$ and $s_3$; but $s_2$ fails to deliver real non-vanishing expectation values for real state vectors.
It is straightforward to show that it is impossible to construct a non-trivial real algebra of observables
on $\mathcal{H}^3_R$ which satisfies the 'uncertainty relations' of the angular momentum algebra.
A real 3-vector can only be interpreted as a classical spin.\\

In the real quantum case, one is left with state vectors whose evolution in time is governed by the propagator
(\ref{blockdiagonalO}) for $n=3$,
i.e. choosing an appropriate basis in $\mathcal{H}^3_R$, the coordinate vector of a state vector evolves
according to
\begin{equation}
\left(\begin{array}{c}
\psi_1 (\tau)\\
\psi_2 (\tau)\\
\psi_3 (\tau)
\end{array}\right) =
\left(\begin{array}{ccc}
\cos (\omega \tau) & \sin (\omega \tau) & 0 \\
- \sin ( \omega \tau) & \cos ( \omega \tau)  & 1 \\
0 & 0 & 1 
\end{array}\right)
\left(\begin{array}{c}
\psi_1 (0)\\
\psi_2 (0)\\
\psi_3 (0)
\end{array}\right)
\end{equation}
with $\omega \in \mathds{R}$. The state space is spanned by a 'vacuum state' and an additional two-dimensional
sector containing a system with 'energy' $\omega$ (or it is spanned by three vacuum vectors in the trivial case $\omega=0$).

%%%%%%%%%%%%%%%%%%%%%%%%%%%%%%%%%%%%%%%%%%%%%%%%%
%%%%%%%%%%%%%%%%%%%%%%%%%%%%%%%%%%%%%%%%%%%%%%%%%

\section{Observables and Measurements in Quantum Mechanics}
It is a widespread misconception in literature that observables in quantum mechanics are
self-adjoint operators due to their real spectrum and the fact that measured data correspond to
real numbers. However, that is not really the point, since there are matrices with real eigenvalues
only which are not hermitian, as the following simple example with eigenvalues $\pm1$ readily demonstrates
\begin{equation}
\left(\begin{array}{cc}
0 & 2 \\
\frac{1}{2} & 0
\end{array} \right)
\left(\begin{array}{c}
1 \\
\pm \frac{1}{2} 
\end{array} \right)
=
\pm
\left(\begin{array}{c}
1 \\
\pm \frac{1}{2} 
\end{array} \right) \, .
\end{equation}
In order to shed some light on the nature of measurements and observables in ordinary {\emph{complex}}
quantum mechanics, a measuring apparatus $\mathcal{A}$ which is capable to detect whether
a quantum mechanical system $\mathcal{S}$ (with state vectors in a Hilbert space $\mathcal{H}_{\mathcal{S}}$)
 is in a state\footnote{States are represented by state {\emph{vectors}}. We will not strictly distinguish the two notions in the
following unless necessary.}
$\mid \uparrow \rangle$ or $\mid \downarrow \rangle \in \mathcal{H}_{\mathcal{S}}$ is considered in
the following.
Of course, using suggestive arrows to label the states $\mid \uparrow \rangle$ and $\mid \downarrow \rangle$
does not imply that we are necessarily dealing with spin states; they can be of a different type.\\

Such states must be stable to a good approximation under an ideal measurement process;
if $\mathcal{S}$ is in a state $\mid \uparrow \rangle$ already before or after a measurement,
then this fact can be verified repeatedly as long as the system and the apparatus are not disturbed too much
in the course of measurements. Otherwise, the measurements would not make much sense.
Both the system $\mathcal{S}$ and the apparatus $\mathcal{A}$
are subsystems of a universe $\mathcal{U}$. The state vectors of $\mathcal{A}$ are elements of a Hilbert space
$\mathcal{H}_{\mathcal{A}}$, and the composition of $\mathcal{S}$ and $\mathcal{A}$ leads to the
completed tensor product space $\mathcal{H}_{\mathcal{S} , \mathcal{A}} = \mathcal{H}_\mathcal{S} \otimes
\mathcal{H}_\mathcal{A}$.\\

An initial state of the apparatus $\mathcal{A}$ ready for measurement is $| A_0 \rangle \in \mathcal{H}_{\mathcal{A}}$;
during the measurement, the dynamics governed by the Hamiltonian $H_{\mathcal{S},\mathcal{A}}$ leads
after some short measuring time to the unitary evolution
(with  $\langle \uparrow \mid \uparrow \rangle_\mathcal{S} =
\langle \downarrow \mid \downarrow \rangle_\mathcal{S} =
\langle A_0 | A_0 \rangle_\mathcal{A} = \langle A_\downarrow | A_\downarrow \rangle_\mathcal{A} =
\langle A_\uparrow | A_\uparrow \rangle_\mathcal{A}  = 1$)
\begin{equation}
\mid\downarrow \rangle | A_0 \rangle \overset{\mathcal{H}_{\mathcal{S},\mathcal{A}}}{ \, \, \longrightarrow \, \,}
\mid\downarrow \rangle | A_\downarrow \rangle  \in \mathcal{H}_{\mathcal{S},\mathcal{A}}
\, , \, \, \,
\mid\uparrow \rangle | A_0 \rangle \overset{\mathcal{H}_{\mathcal{S},\mathcal{A}}}{\, \, \longrightarrow \, \,}
\mid\uparrow \rangle | A_\uparrow \rangle \in \mathcal{H}_{\mathcal{S},\mathcal{A}} =
\mathcal{H}_{\mathcal{S}} \otimes \mathcal{H}_{\mathcal{A}} \, .
\label{unitary_time_evolution}
\end{equation}
Unitarity of the time evolution requires conservation of scalar products, hence
\begin{equation}
\langle \uparrow \mid \downarrow \rangle_{\mathcal{S}} \langle A_0 | A_0 \rangle_{\mathcal{A}} =
\langle \uparrow \mid \downarrow \rangle_{\mathcal{S}} \langle A_\uparrow | A_\downarrow
\rangle_{\mathcal{A}} \, .
\label{measurement_overlap}
\end{equation}
Equation (\ref{measurement_overlap}) is crucial. One might continue with a mistake and divide it
by $\langle \uparrow \mid \downarrow \rangle_{\mathcal{S}}$ and ending up this way with
$\langle A_0 | A_0 \rangle_{\mathcal{A}} = \langle A_\uparrow | A_\downarrow \rangle_{\mathcal{A}} =1$.
But this equation would imply $| A_\downarrow \rangle = | A_\uparrow \rangle$, and the apparatus $\mathcal{A}$
would be incapable to distinguish between the two states $\mid\uparrow \rangle$ und $\mid \downarrow \rangle$.\\

{\emph{Only}} when $\langle \downarrow \mid \uparrow \rangle_{\mathcal{S}} =0$ holds true, equation
(\ref{measurement_overlap}) makes sense \cite{Zurek}. One finally concludes that states can be discriminated by an
apparatus only when they are mutually orthogonal, and requiring that a measuring device delivers real numbers
implies that the corresponding observable is a self-adjoint operator on $\mathcal{H}_\mathcal{S}$.

%%%%%%%%%%%%%%%%%%%%%%%%%%%%%%%%%%%%%%%%%%%%%%%%%
%%%%%%%%%%%%%%%%%%%%%%%%%%%%%%%%%%%%%%%%%%%%%%%%%

\section{Conclusions}
Historically, Erwin Schr\"odinger has been struggling between choosing real or complex wave functions in
the early days of quantum mechanics. His concerns about using the imaginary number $i=\sqrt{-1}$
can be sensed in one of his letters to Hendrik Antoon Lorentz \cite{Letter1}:  ``What is unpleasant here, and indeed directly
to be objected to, is the use of complex numbers.  $\psi$ is surely fundamentally a real function~$\dotsc$.''
Few days later, he wrote to Max Planck \cite{Letter2}: 
``The time dependence must be given by $\psi \sim P.R. \, \left(e^{\pm\frac{2\pi iEt}{h}}\right)$,''
(where $P.R.$ ({\emph{pars realis}}) denotes the real part) and then:
``or, what is the same thing, we must have $\frac{\partial^{2}\psi}{\partial t^2} = -\frac{4\pi^{2}E^2}{h^2} \psi$.''
So Schr\"odinger clearly expressed his initial desire to avoid complex numbers by differentiating his wave function
$\psi$ twice.\\

Shortly after Schr\"odinger's letters to Lorentz and Planck, he wrote in one of his famous papers which introduced
quantum wave mechanics to physics (translated from \cite{Schroedinger}):
``We will require the complex wave function $\psi$ to satisfy one of these two equations. 
Since the conjugate complex function $\bar{\psi}$ will then satisfy the
{\emph{other}} equation, we may take the real part of $\psi$ as the {\emph{real wave function}}
(if we require it).'' He concludes at the end of the paper:
``Meantime, there is no doubt a certain severity in the use of a {\emph{complex}} wave function. 
If it were unavoidable {\emph{in principle}}, and not merely a facilitation of the calculation, this would
mean that there are in principle {\emph{two}} wave functions, which must be used {\emph{together}}
in order to obtain information on the state of the system.  This somewhat unpleasant conclusion admits, I believe,
of the very much more congenial interpretation that the state of the system is given by a real function and its
derivative with respect to time.''\\

Non-trivial orthogonal dynamics of a quantum system described within in a real $n=2m$-dimensional Hilbert space
$\mathcal{H}_R^n$ automatically leads to the emergence of physically equivalent pairs of orthogonal states
combining to rotating real 'bi-state vectors' which span invariant two-dimensional subspaces in $\mathcal{H}_R^n$.
These bi-state vectors exhibit an analogous temporal behavior under $SO(2)$ transformations to the single eigenstate
vectors of a Hamiltonian in complex (unitary) quantum mechanics which oscillate under $U(1)$ transformations.\\

One should keep in mind that Hilbert space elements are viewed as {\emph{universal}} wave functions in this paper,
and no statements have been made about a physical interpretation of the eigenstates of the universal Hamiltonian
in the complex case or the two-dimensional eigenspaces of the universal Hamiltonian in the real case.
However, considering a quantum mechanical subsystem of a real quantum universe and trying to induce a reduction of the
state on a real bi-state by measurement will only change the phase of this system, which will then restart to oscillate.
The frequency of the electron clock  at rest $\omega_e = m_e c^2 / \hbar =
c/ \mkern0.75mu\mathchar '26\mkern -9.75mu\lambda =  7.7634 \cdot 10^{20} \mbox{s}^{-1}$
is huge from an experimental point of view despite the small mass $m_e$ of the particle. 
Therefore, there might be a physical origin of the quantum mechanical phase which, however, is hard to observe
due to its dynamic behavior (see also comments in \cite{Hestenes1, Hestenes2}).\\

The generalization of the present dicussion to {\emph{separable}} infinite dimensional Hilbert spaces
should not pose severe technical problems.

%%%%%%%%%%%%%%%%%%%%%%%%%%%%%%%%%%%%%%%%%%%%%%%%%
%%%%%%%%%%%%%%%%%%%%%%%%%%%%%%%%%%%%%%%%%%%%%%%%%

\end{document}